\documentstyle[aps,twocolumn,floats,prd,psfig]{revtex}

\newcommand{\mybib}[2]{\bibitem{#2}}

\newcommand{\ApJ}{Astrophys. J.}
\newcommand{\PRL}{Phys. Rev. Lett.}
\newcommand{\PRD}{Phys. Rev. D}
\newcommand{\MNRAS}{MNRAS}
\newcommand{\AsAs}{A\&A}
\newcommand{\aut}[2]{{#2.\ #1}}
\newcommand{\refs}[6]{#2, {\bf #3} {#4} (#5)}
\newcommand{\urefs}[4]{#2, #3 (#4)}
\newcommand{\brefs}[4]{{\it #1} (#2, #3, #4)}
\newcommand{\amp}{and }

\newcommand{\bn}{\hat{\bf n}}

\newcommand{\bk}{{\bf k}}
\newcommand{\bx}{{\bf x}}
\newcommand{\prim}{{p}}
\newcommand{\len}{{\phi}}
\newcommand{\del}{{d}}

\newcommand{\intr}{{\rm i}}

\newcommand{\dC}{T}
\newcommand{\dE}{E}
\newcommand{\dB}{B}
\newcommand{\dX}{X}
\newcommand{\rms}{{\rm rms}}

\newlength{\tskip}
\newlength{\colwidth}

\newcommand{\beq}{\begin{equation}}
\newcommand{\eeq}{\end{equation}}
\newcommand{\beqa}{\begin{eqnarray}}
\newcommand{\eeqa}{\end{eqnarray}}

\def\simlt{\lesssim}
\def\simgt{\gtrsim}
\newcommand{\wj}{\left(
                          \begin{array}{ccc}
                          l  &  l_1  & l_2 \\
                            0  &  0    &  0
                          \end{array}
                          \right)}
\newcommand{\wjma}[6]{\left(
                           \begin{array}{ccc}
         #1 & #2  & #3  \\
         #4 & #5  & #6
                           \end{array}
                   \right)}

\begin{document}
\twocolumn[\hsize\textwidth\columnwidth\hsize\csname
@twocolumnfalse\endcsname

\title{Gravitational Time Delay Effects on CMB
Anisotropies}
\author{Wayne Hu$^{1,2}$, Asantha Cooray$^2$}
\address{
$^1$Institute for Advanced Study, Princeton, NJ 08540\\
$^2$Department of Astronomy and Astrophysics, University of Chicago,
Chicago IL 60637\\
E-mail:  whu@ias.edu, asante@hyde.uchicago.edu}

\maketitle

\begin{abstract}
We study the effect of gravitational time delay on the power spectra
and bispectra of the cosmic microwave background (CMB) temperature and
polarization anisotropies.   The time delay effect modulates the spatial 
surface at recombination on which temperature anisotropies are observed, 
typically by $\sim$ 1 Mpc.  While this is a relatively large 
shift, its observable effects in the temperature and polarization fields 
are suppressed by geometric considerations.  The leading order effect is 
from its correlation with the closely related gravitational lensing effect.
The change to the temperature-polarization cross power spectrum is of
order $0.1\%$ and is hence comparable to the cosmic variance for
the power in the multipoles around $\ell \sim 1000$. While unlikely
to be extracted from the data in its own right, its  omission
in modeling would produce a systematic error comparable to this limiting
statistical error and, in principle, is relevant
for future high precision experiments.  Contributions to the
bispectra result mainly from correlations with the Sachs-Wolfe effect
and may safely be neglected in a low density universe.
\end{abstract}
\vskip 0.5truecm
]


\section{Introduction}

In order that the full potential of anisotropies in the cosmic
microwave background (CMB) temperature and polarization fields be realized,
effects that have been previously dismissed as negligible in their
own right must now be reconsidered as potential sources of systematic
error.  Projections as to the ability of CMB experiments to measure
fundamental cosmological quantities \cite{est} precisely
and reveal information about the structure
formation process from secondary effects
\cite{sec} rely on the fact that statistical errors from the sampling of
a finite sky rapidly decrease toward smaller angular scales.
Statistical errors in the power spectra decline from $\sim 1\%$ at degree
scales to $\sim 0.1\%$ at the several arcminute scale.
To achieve this precision in practice, all physical, astrophysical
and instrumental effects at this level
must be included in the analysis to avoid generating systematic errors
that are comparable to the statistical errors.

A host of physical effects contribute to the anisotropies at 
second order in perturbation theory \cite{HuScoSil94,DodJub95,PynCar96,MatMolBru98}.
Since primary anisotropies are formed at
recombination when the cosmological density perturbations
are at the $10^{-5}$
level, most second order effects are entirely negligible.  There are
two general ways in which higher order effects can be important.
Firstly, the primordial perturbations responsible for the primary
anisotropies grow into non-linear structures today by gravitational
instability.   Effects that take advantage of this fact mainly involve
scattering of CMB photons at low redshifts in large-scale structure
and non-linear objects \cite{SunZel70,Vis87,GruHu98,Hu00b}.
Secondly, since recombination, CMB photons propagate across
essentially the whole horizon volume.
Intrinsically small effects can accumulate along the path.  Indeed
it is well known that the gravitational lensing of CMB photons has
a substantial effect on the power spectrum of the anisotropies
\cite{Sel96,ZalSel98}.

In addition to the lensing effect, gravitational potentials
of large-scale structure  contributes a time-delay \cite{Sha64}
that accumulates along the path -- an effect familiar
from studies of the light-curves of lensed quasars
(e.g. \cite{SchEhlFal92}).   In the case of the CMB,
the time-delay warps the {\it spatial} surface at recombination from which
the primary anisotropies arise \cite{PynCar96}.
Because the lensing depends on the angular gradient of the projected
potentials whereas the delay depends on the projected potential itself,
the fractional delay is generically smaller than lensing and
has not been explicitly calculated in the literature.
We shall see however that because of the angular smoothness
(or coherence) of the lensing, the reduction in amplitude is not
in and of itself large.  Furthermore, gravitational time
delays are strongly correlated
with lensing, leading to additional effects in the power spectrum.
Indeed, the typical perturbation in comoving units is on the order of
$1$ Mpc.  The effect of gravitational time-delay on the spectra of
temperature and polarization anisotropies therefore merits further
study.

In \S \ref{sec:formalism} we present the formalism required
to understand these gravitational effects on the temperature
and polarization fields.  We proceed in
\S \ref{sec:power} and \ref{sec:polarization} to evaluate
the delay and lensing-delay correlation effects on the
power spectra of temperature and polarization anisotropies
respectively.  In \S \ref{sec:bispectrum}, we consider their
effects on the bispectra (three point correlations).  We
conclude in \S \ref{sec:discussion} with a discussion of our
results.

To illustrate our calculations, we
 assume a cold dark matter model (CDM) with a cosmological constant ($\Lambda$) with parameters $\Omega_c=0.30$ for the CDM density,
$\Omega_b=0.05$ for the baryon density, $\Omega_\Lambda=0.65$ for the
vacuum density, $h=0.65$ for the dimensionless Hubble
constant and a scale invariant spectrum of
primordial fluctuations, normalized to COBE \cite{BunWhi97}.

\begin{figure}[t]
\centerline{\psfig{file=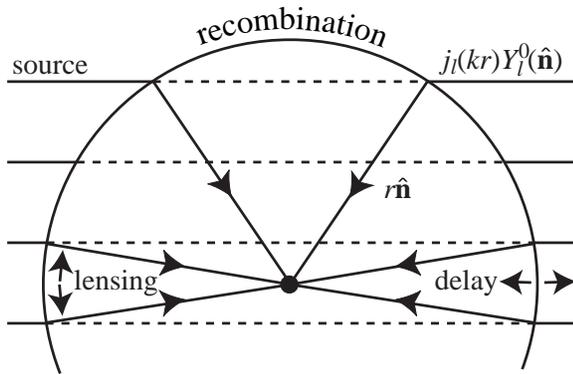,width=3.0in}}
\caption{Gravitational Lensing vs. Time Delay.  Lensing introduces an angular
perturbation in the mapping of a plane-wave source field at recombination
onto anisotropies today.  Time delay introduces a radial modulation.
When the wavevector is perpendicular to the line-of-sight, features
in the angular spectrum -- such as the acoustic peaks --
are created, geometrically distinguishing the otherwise similar
lensing and delay effects.
}
\label{fig:delay}
\end{figure}

\section{Formalism}
\label{sec:formalism}

As the CMB photons propagate to the observer from the recombination
epoch ($z \sim 10^3$) through large-scale structure in the universe,
they suffer the effects of gravitational
lensing and time delay.  These effects are both formally second order
in perturbation theory because they would leave a homogeneous and isotropic
CMB unperturbed (c.f.~\cite{CheWuJia00}).  There are a host of other second order effects
\cite{HuScoSil94,DodJub95,PynCar96}.  To the extent that they
are uncorrelated with each other, they may be viewed as independent effects.
The lensing and time delay effects are strongly correlated because
both arise from the gravitational potentials
of large scale structure and must be considered together.  We therefore
begin with a review of lensing effects.

\subsection{Lensing}
\begin{figure}[t]
\centerline{\psfig{file=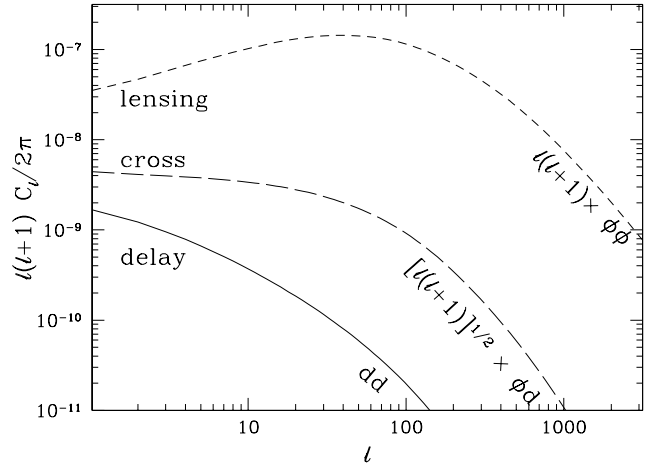,width=3.5in}}
\caption{Power spectra for the lensing deflection angles ($\len\len$),
time-delay ($\del\del$) and deflection-delay
cross correlation ($\len\del$).
The underlying lensing potential spectrum $C_\ell^{\len\len}$ and
cross spectrum $C_\ell^{\len\del}$ are weighted by $\ell(\ell+1)$
and $[\ell(\ell+1)]^{1/2}$ respectively to reflect the angular gradients
in the deflection angles.}
\label{fig:grav}
\end{figure}
Lensing involves a deflection that remaps the temperature and polarization
fields according to the angular gradient of the lensing-weighted projected
potential $\bn \rightarrow \bn + \nabla \phi(\bn)$ (see Fig.~\ref{fig:delay})
\begin{equation}
    \phi(\bn) = -2 \int d\eta g_{\phi}(\eta) \Phi(r\bn,\eta)\,,
\label{eqn:projectedpot}
\end{equation}
where
\begin{equation}
g_{\phi}(\eta) = {1 \over r} \int_0^\eta d\eta'  \,
	        \dot\tau e^{-\tau}
                {r'-r \over r'} \,.
\label{eqn:lensingeff}
\end{equation}
Here overdots represent derivatives with respect to conformal time
$\eta = \int dt/a$ and $\Phi(\bx,\eta)$ is the gravitational potential.
In an open universe, the conformal distance traveled by
a photon $r(\eta) = \eta_0-\eta$ should be replaced by angular diameter distances.
We take comoving units where $c=1$ throughout.
The projected potential itself is a field on the sky and may be
decomposed into multipole moments
\begin{eqnarray}
\len(\bn) = \sum_{\ell m} \len_{\ell m} Y_\ell^m (\bn)  \,,
\end{eqnarray}
and described by its power spectrum
\begin{eqnarray}
\left< \len_{\ell m}^* \len_{\ell' m'} \right> = \delta_{\ell,\ell'}
	\delta_{m,m'} C_\ell^{\len\len} \,.
\end{eqnarray}
This power spectrum is shown in Fig.~\ref{fig:grav}.

A useful measure of the amplitude of the lensing effects is the rms
deflection angle $\theta_{\rms}^2$ as defined by
\begin{equation}
\theta_{\rms}^2 = \sum_{\ell=1}^\infty {2 \ell+1 \over 4\pi} \ell(\ell+1) C_\ell^{\len\len}\,.
\label{eqn:rmslensing}
\end{equation}
The factor $\ell(\ell+1)$ reflects the angular gradient in the definition
of the deflection.
Note that the $\ell=0$ monopole does not contribute since
its angular gradient vanishes.
In the fiducial $\Lambda$CDM model $\theta_{\rms}=7.5 \times 10^{-4}$ or $2.6'$.
Note that this is in sharp contrast with the angular {\it coherence} of the
deflection angle.  The variance
$\theta_{\rms}^2$ reaches half its total value by $\ell_{1/2}= 30$ in the fiducial
model or an angle of $\theta_{1/2} \equiv 2\pi/\ell_{1/2}\approx 0.02$ ($1^\circ$).
The smaller scale potential fluctuations tend to produce deflections that cancel
out along the line of sight.

\subsection{Time Delay}

Now consider the time-delay effect.  Photons follow null geodesics in the
perturbed metric so that in the fixed time interval since last scattering, the
distance traveled by the photons is perturbed as $r \rightarrow
r [1 + d(\bn)]$ (see Fig.~\ref{fig:delay}) where
\begin{equation}
    d(\bn) = -{2 \over \eta_0} \int d\eta e^{-\tau} \Phi(r\bn,\eta)\,.
\label{eqn:projectedtd}
\end{equation}
This is referred to in the literature as the potential or Shapiro
time delay.   The geometric time delay is of order the deflection angle
squared and is hence substantially smaller for time-delays across
angular scales much larger than $\theta_{\rms}$.

The delay field may also be expanded in spherical harmonics
\begin{eqnarray}
\del(\bn) &=& \sum_{\ell m} \del_{\ell m} Y_\ell^m (\bn) \,,
\end{eqnarray}
and characterized by a power spectrum
\begin{eqnarray}
\left< \del_{\ell m}^* \del_{\ell' m'} \right> &=& \delta_{\ell,\ell'}
	\delta_{m,m'} C_\ell^{\del\del}  \,.
\end{eqnarray}
Although both the
lensing and time-delay effects are based on the gravitational potential
projected along the line of sight, there is an important difference between the
two.  Lensing depends on the angular gradient of the potential and hence its
observable consequences are weighted by $\ell(\ell+1)$.  This has the effect
of increasing the magnitude of the effects and weighting it to higher multipoles.

We can see these effects in the rms delay \cite{caveat}
\begin{equation}
d_{\rms}^2 = \sum_{\ell=1}^\infty {2 \ell+1 \over 4\pi} C_\ell^{\del\del}\,.
\label{eqn:rmsdelay}
\end{equation}
For the power spectrum of the fiducial $\Lambda$CDM model (see Fig.~\ref{fig:grav})
$d_{\rms}= 5.4 \times 10^{-5}$ or $\sim 0.5 h^{-1}$ Mpc.  Although this is a
relatively large shift, we shall see that its observable consequences are
reduced due to the large coherence scale of the delay, $\ell_{1/2}=2$.

Given these considerations as to the magnitude and coherence scale of the effects,
delay effects can be enhanced through their cross-correlation with
lensing
\begin{eqnarray}
\left< \len_{\ell m}^* \del_{\ell' m'} \right> &=& \delta_{\ell,\ell'}
	\delta_{m,m'} C_\ell^{\len\del}  \,.
\end{eqnarray}
The two fields tend to be well correlated
as can be seen by the rms
\begin{equation}
c_{\rms}^2 = \sum_{\ell=1}^\infty {2 \ell+1 \over 4\pi} \sqrt{\ell(\ell+1)}
	C_\ell^{\len\del}\,,
\end{equation}
where $c_{\rms}= 1.2 \times 10^{-4}$ and $\ell_{1/2}=5$ in the fiducial
$\Lambda$CDM model (see Fig.~\ref{fig:grav}).
To evaluate the observable consequences, we now turn to the angular
and spatial structure of the temperature and polarization fields.

\begin{figure}[t]
\centerline{\psfig{file=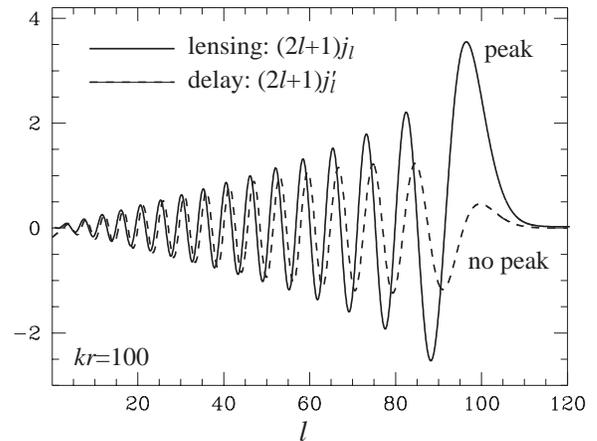,width=3.0in}}
\caption{Projection functions for the lensing and delay effects.
Lensing involves $j_\ell(kr)$
whose strong peak in $\ell$ can retain source features; the $j_\ell'(kr)$
of the time-delay cannot.  The product of these functions reflects
the cross correlation and is suppressed due to the phase difference.
}
\label{fig:jl}
\end{figure}

\subsection{Temperature Field}
\label{sec:temp}

The CMB temperature field on the sky may be written
implicitly as the projection of
sources $S$ which contribute in the optically thin regime and are so weighted
by $e^{-\tau}$ where $\tau$ is the optical depth (see Fig.~\ref{fig:delay}).
In general, these
sources have intrinsic angular structure of their own and are characterized by the
spherical harmonic moments of their Fourier amplitude $S_{\ell_\intr}^{m_\intr}(k)$.
Explicit forms for the sources are given in \cite{HuWhi97}.

The contribution from a given wavenumber $k$ to the temperature field on the sky
today may be formally expressed as
\begin{equation}
\Theta(\bn;\bk) = \int_0^{\eta_0} d\eta e^{-\tau} \sum_{\ell_\intr m_\intr}
                   S_{\ell_\intr}^{m_\intr} (\eta;k) G_{\ell_\intr}^{m_\intr}(r\bn;\bk)\,,
\end{equation}
where $r \equiv \eta_0-\eta$,
\begin{equation}
G_\ell^m(\bx;\bk) = (-i)^{\ell} \sqrt{ 4\pi \over 2\ell+1} Y_\ell^m (\bn) \exp(i \bk \cdot \bx) \,.
\end{equation}
In an open universe, the plane waves must be replaced with eigenfunctions of the
Laplacian in a curved space.
Note that we will often omit the $k$-index where no confusion will arise.

The angular structure of these relations can be simplified by considering
a specific frame where ${\bf z} \parallel \bk$ and
\begin{equation}
\exp(i \bk \cdot \bx) = \sum (-i)^\ell \sqrt{4\pi(2\ell+1)} j_\ell(kr) Y_\ell^0(\bn) \,.
\label{eqn:plane}
\end{equation}
Provided that the angular basis does not change in transit, one
can then sum up the orbital angular momentum from the plane wave with the intrinsic
angular momentum of the source.   This assumption is the angular equivalent of
the ``Born approximation" where the lensing is evaluated on unperturbed trajectories;
we shall see in the next section
that it is a good approximation for the lensing and time-delay effects on sources
with low order intrinsic angular structure.

In this special basis, the product of the
intrinsic ($Y_{\ell_\intr}^{m_\intr}$)
and plane wave ($Y_{\ell}^0$) angular momentum may be reexpressed
through the addition of angular momentum.
The temperature field then becomes
\cite{HuWhi97}
\begin{eqnarray}
\Theta(\bn;\bk) & \equiv &
	\sum_{\ell m} \Theta_{\ell m}(k) Y_\ell^{m}(\bn) \nonumber\\
&=&	\sum_{\ell m_\intr}
	I_{m_\intr} [j_\ell] Y_\ell^{m_\intr} (\bn) \,.
\label{eqn:tempmoments}
\end{eqnarray}
Here, the operator
\begin{eqnarray}
I_{m_\intr}[j_\ell] &\equiv&
		\int_0^{\eta_0} d\eta e^{-\tau}
		 \sqrt{4\pi(2\ell+1)}  \nonumber\\
		&& \quad\times
			\sum_{\ell_\intr}
	S_{\ell_\intr}^{m_\intr} (\eta;k) j_\ell^{\ell_\intr m_\intr}(kr)\,,
\label{eqn:operator}
\end{eqnarray}
and $j_\ell^{\ell_\intr m_\intr}$ are linear combinations of $j_\ell$ with weights
given by the Clebsch-Gordon coefficients of the coupling.
The fundamental functions are $j_\ell^{00}=j_\ell$ (for isotropic
perturbations, e.g. gravitational potentials) and
\begin{eqnarray}
j_\ell^{22} &=& \sqrt{{3 \over 8}{(\ell+2)! \over (\ell-2)!}}{1 \over 2\ell+1}\left[
		{j_{\ell-2}+j_\ell \over (2\ell-1)}
		+
		      {j_\ell +j_{\ell+2} \over (2\ell+3)}\right] \,,
\label{eqn:tensorjl}
\end{eqnarray}
(for transverse quadrupole sources, e.g. gravitational waves).
Note $j_\ell^{22} \propto j_\ell(x)/x^2$; we have written it out
here to emphasize linearity. Others are given in \cite{HuWhi97},
but can be written in terms of these fundamental functions through
integration by parts \cite{SelZal96}.

Since the basis for the expansion is linked to the direction of $\bk$,
integrating over modes to obtain the final power spectrum in principle requires a series
of rotations into a fixed basis.  In practice, the statistical
homogeneity of the source
and isotropy of the angular distribution requires that the $k$-modes add in
quadrature and the angular power spectrum is independent of $m$ so that
we may replace individual multipoles with an average over $m$
\begin{eqnarray}
C_\ell^{\Theta\Theta} &=& \int {d k \over k} {k^3 \over 2\pi^2} {1 \over 2 \ell+1} \sum_{m}
	\left< \Theta_{\ell m}^*(k) \Theta_{\ell m}(k) \right> \,.
\end{eqnarray}
For the primary anisotropies, the power spectrum is simply $C_\ell^{\Theta\Theta} = T_\ell^{00}$, where
\begin{eqnarray}
T_{\ell}^{a b}   & = &
	\int {d k \over k} {k^3 \over 2\pi^2} {(k\eta_0)^{a + b}
	\over 2 \ell+1} \sum_{m_\intr} I_{m_\intr}^* [j_\ell^{(a)}]
	I_{m_\intr} [j_\ell^{(b)}]\,.
\label{eqn:tdefn}
\end{eqnarray}
Here we have used the short hand convention that $j_\ell^{(0)} \equiv
j_\ell$, $j_\ell^{(1)} \equiv j_\ell'$ and $j_\ell^{(2)} \equiv j_\ell''$
with primes denoting derivatives with respect to the argument.

\begin{figure}[t]
\centerline{\psfig{file=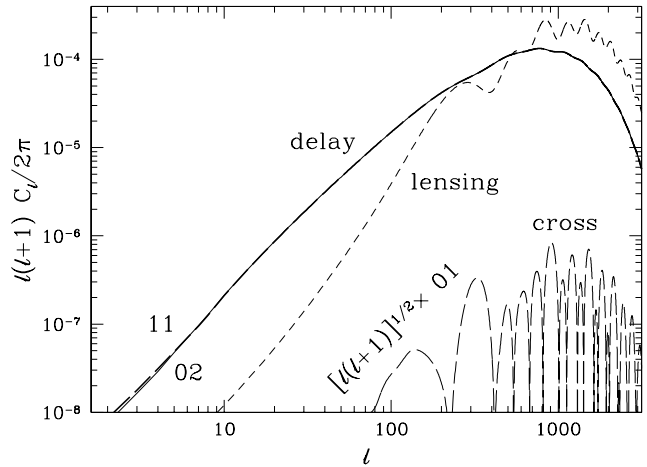,width=3.5in}}
\caption{Power spectra for the angular ($\ell(\ell+1) T_\ell^{00}$), radial
($T_\ell^{11}$) and angular-radial cross
gradients $(\sqrt{\ell(\ell+1)} T_\ell^{01}) $ of the primary anisotropies.
The cross gradient spectrum has been weighted to reflect the $\ell$ contributions
from the angular gradient.}
\label{fig:gradpower}
\end{figure}

For the higher order lensing and delay effects, we expand
equation~(\ref{eqn:tempmoments}) for the temperature field to second
order in the relevant quantities
\begin{eqnarray}
\Theta(\bn) & = &            \Theta^0(\bn)
		 +           \Theta^{\len}(\bn)
		 +           \Theta^{\del}(\bn)  \nonumber\\
	&&\quad	+{1 \over 2} \Theta^{\len^2}(\bn)
                +            \Theta^{\len\del}(\bn)
		+{1 \over 2} \Theta^{\del^2}(\bn) \,,
\label{eqn:expand}
\end{eqnarray}
where $\Theta^0 (\bn)$ is the zeroth order contribution from the primary anisotropies,
\begin{eqnarray}
\label{eqn:pert}
\Theta^{\len}(\bn) & = &  \sum_{\ell m_\intr} I_{m_\intr}[j_\ell] \,
			\nabla_i \phi(\bn)\,  \nabla^i Y_\ell^{m_\intr}(\bn) \,,\nonumber\\
\Theta^{\del}(\bn) & = &  \sum_{\ell m_\intr} I_{m_\intr}[j_\ell'] \, (k\eta_0) \, d(\bn) \, Y_\ell^{m_\intr}(\bn) \,,\\
\Theta^{\len^2}(\bn) & = &  \sum_{\ell m_\intr} I_{m_\intr}[j_\ell] \,
			\nabla_i \phi(\bn)\,  \nabla_j \phi(\bn) \,
			\nabla^i \nabla^j Y_\ell^{m_\intr}(\bn) \,,\nonumber\\
\Theta^{\len\del}(\bn) &=& \sum_{\ell m_\intr} I_{m_\intr}[j_\ell']\, (k\eta_0) \,d(\bn)\, \nabla_i \phi(\bn)\, \nabla^i
			Y_\ell^{m_\intr}(\bn)\,, \nonumber\\
\Theta^{\del^2}(\bn) & = &  \sum_{\ell m_\intr} I_{m_\intr}[j_\ell''] \, (k\eta_0)^2 \, d(\bn) \,d(\bn) \,
			Y_\ell^{m_\intr}(\bn) \,.\nonumber
\end{eqnarray}
We shall see that the evaluation of these effects reduces to the computation
of the higher order  derivative power spectra in equation~(\ref{eqn:tdefn}).
These power spectra are easily evaluated with minimal modifications to the publically
available CMBFAST code: we simply replace
$j_\ell$ with $j_\ell'$ and $j_\ell''$ and leave the evolution and
integration over the sources unchanged.
To the extent that the underlying sources themselves are smooth,
integration by parts on equation~(\ref{eqn:tdefn}) shows that $T_\ell^{(a \pm 1)(b\mp 1)}
\approx -T_\ell^{a b}$, e.g. $T_\ell^{11} \approx -T_\ell^{02}
= -T_\ell^{20}$.
This also implies that terms such as $T_\ell^{01} = T_\ell^{10}$ are
suppressed; mathematically this is due to the lack of correlation between
$j_\ell$ and $j_\ell'$ (see Fig.~\ref{fig:jl}).
These spectra are compared in Fig.~\ref{fig:gradpower}.

There is an additional effect from lensing due to the fact that sources are in general
anisotropic and gravitational lensing changes the angle at which they are viewed.
This formally violates the assumption that the orbital and intrinsic angular momentum
of sources can simply be added without additional remapping.
The anisotropy in the actual sources, however,
are confined to $\ell \le 2$.  To obtain an order unity effect, the deflections
must change the viewing angle by order unity in radians.
To obtain an order unity effect for lensing, the deflections
need change the viewing direction by
only a wavelength of the angular perturbation $\ell^{-1}$.  Thus
at sufficiently high observed $\ell$, deflections
always win.

\subsection{Polarization Field}

The complex Stokes parameter $P_\pm = Q \pm iU$ of the polarization on the sky
can be expressed implicitly as the projection
of the quadrupole moments of the photon temperature and polarization
distributions at last scattering
\begin{eqnarray}
P_\pm (\bn; \bk) &=& \int_0^{\eta_0} d\eta e^{-\tau}
		\sum_{m_\intr} Q^{m_\intr}(\eta;k)\,
		{}_{\pm 2}G_{2}^{m_\intr}(r \bn;\bk)\,,
\end{eqnarray}
where $Q^{m_i} = -\dot\tau (\sqrt{6}\Theta_{2 m} - 6 E_{2 m})/10$ (see
\cite{HuWhi97}, for an explicit definition)
and
\begin{eqnarray}
{}_{\pm 2} G_2^m(\bx;\bk) = (-i)^{\ell} \sqrt{ 4\pi \over 2\ell+1} {}_{\pm 2} Y_2^m
(\bn) \exp(i \bk \cdot \bx) \,.
\end{eqnarray}
Here, ${}_{\pm 2} Y_\ell^m$ are the spin-2 spherical harmonics \cite{Goletal67}.
A recoupling of the spin spherical harmonics as in equation~(\ref{eqn:plane}) yields
\begin{eqnarray}
P_\pm (\bn; \bk) &=& \sum_{\ell m} [E_{\ell m}(k) \pm i B_{\ell m}(k)] \, {}_{\pm 2} Y_\ell^m(\bn)
\nonumber\\
		      &=& \sum_{\ell m_\intr}
			P_{m_\intr}[{}_\pm \alpha_\ell]
			{}_{\pm 2} Y_\ell^{m_\intr}(\bn)\,,
\end{eqnarray}
where the operator
\begin{eqnarray}
P_{m_\intr}[{}_\pm \alpha_\ell] &\equiv &
		\int_0^{\eta_0} d\eta e^{-\tau}
		 \sqrt{4\pi(2\ell+1)}  \nonumber\\
&& \quad\times
	\sum_{\ell_\intr}
Q^{m_\intr} (\eta;k) {}_\pm \alpha_\ell^{m_\intr}(kr)\,.
\label{eqn:poperator}
\end{eqnarray}
Here, ${}_\pm \alpha_\ell^{m_\intr} = \epsilon_\ell^{m_\intr} \pm i
\beta_\ell^{m_\intr}$  where the latter are combinations of $j_\ell$
given in \cite{HuWhi97} that define the projection
of the source onto the
$E$ and $B$ polarization modes.  For quadrupole sources related to density $(m_\intr=0)$ fluctuations,
$\epsilon_\ell^0 = j_\ell^{22}$ (see eqn. [\ref{eqn:tensorjl}]) and
   $\beta_\ell^0=0$ so that there are no contributions to $B$-parity
polarization.

The power spectra for the polarization and temp\-erature-polar\-ization cross
correlation are defined as
\begin{eqnarray}
C_\ell^{F  F'} &=& \int {d k \over k} {k^3 \over 2\pi^2} {1 \over 2 \ell+1} \sum_{m}
	\left< F_{\ell m}^*(k) F_{\ell m}'(k) \right> \,.
\end{eqnarray}
where $F$ and $F'$ take on the values $\Theta$,$E$ and $B$.
For the primary polarization, we define $C_\ell^{EE} = E_\ell^{00}$,
$C_\ell^{BB} = B_\ell^{00}$,
$C_\ell^{\Theta E} = X_\ell^{00}$, where
\begin{eqnarray}
E_\ell^{a b} &=& \int {d k \over k} {k^3 \over 2\pi^2} {(k\eta_0)^{a + b}
		\over 2\ell+1} \sum_{m_\intr}
		P_{m_\intr}^*[\epsilon_\ell^{(a)}]
		P_{m_\intr}[\epsilon_\ell^{(b)}] \,, \nonumber\\
B_\ell^{a b} &=& \int {d k \over k} {k^3 \over 2\pi^2} {(k\eta_0)^{a + b}
		\over 2\ell+1} \sum_{m_\intr}
		P_{m_\intr}^*[\beta_\ell^{(a)}]
		P_{m_\intr}[\beta_\ell^{(b)}] \,, \\
\label{eqn:poldefn}
X_\ell^{a b} &=&
		\int {d k \over k} {k^3 \over 2\pi^2} {(k\eta_0)^{a + b}
		\over 2\ell+1} \sum_{m_\intr}
	I_{m_\intr}^*[j_\ell^{(a)}] P_{m_\intr}[\epsilon_\ell^{(b)}]
		\nonumber\,.
\end{eqnarray}
Other combinations vanish due to parity considerations.
As in equation~(\ref{eqn:tdefn}) the indices $a$ and $b$ refer to derivatives
of the underlying Bessel functions.
The higher order gravitational effects on the polarization field may
be evaluated through a second order expansion of the polarization field
\begin{eqnarray}
P_\pm(\bn) & = &  P_\pm^0(\bn)  + P_\pm^{\len}(\bn)
		                    + P_\pm^{\del}(\bn)
\nonumber\\
           && \quad	+  {1 \over 2}P_\pm^{\len^2}(\bn)
                        +  P_\pm^{\len\del}(\bn)
		        +  {1 \over 2}P_\pm^{\del^2}(\bn)
\,,
\end{eqnarray}
where $P_\pm^0 (\bn)$ is the zeroth order contribution from the primary anisotropies,
\begin{eqnarray}
P_\pm^{\len}(\bn) & = &  \sum_{\ell m_\intr} P_{m_\intr}[{}_\pm \alpha_\ell] \,
			\nabla_i \phi(\bn)\,  \nabla^i
			{}_{\pm 2} Y_\ell^{m_\intr}(\bn) \,,\\
\label{eqn:pperturbed}
P_\pm^{\del}(\bn) & = &  \sum_{\ell m_\intr} P_{m_\intr}[{}_\pm \alpha_\ell']
			\, (k\eta_0) \, d(\bn) \,
			{}_{\pm 2} Y_\ell^{m_\intr}(\bn) \,,\nonumber\\
P_\pm^{\len^2}(\bn) & = &  \sum_{\ell m_\intr} P_{m_\intr}[{}_\pm \alpha_\ell] \,
			\nabla_i \phi(\bn)\,  \nabla_j \phi(\bn) \,
			\nabla^i \nabla^j
			{}_{\pm 2} Y_\ell^{m_\intr}(\bn) \,,\nonumber\\
P_\pm^{\len\del}(\bn)
	            & = &  \sum_{\ell m_\intr} P_{m_\intr}[{}_\pm \alpha_\ell']\,
			(k\eta_0) \,d(\bn)\, \nabla_i \phi(\bn)\, \nabla^i
			{}_{\pm 2} Y_\ell^{m_\intr}(\bn)\,.\nonumber\\
P_\pm^{\del^2}(\bn) & = &  \sum_{\ell m_\intr} P_{m_\intr}[{}_\pm \alpha_\ell'']
			\, (k\eta_0)^2 \, d(\bn) \,d(\bn) \,
			{}_{\pm 2} Y_\ell^{m_\intr}(\bn) \,.\nonumber
\end{eqnarray}
The evaluation of the effects will involve the higher order derivative power
spectra of equation~(\ref{eqn:poldefn}).  Again we modify
CMBFAST to calculate these spectra.  Just as for the temperature
power spectra $E_\ell^{(a\pm 1)(b \mp 1)}
\approx -E_\ell^{a b}$ and similarly for $B$ and $X$.  This
also implies that terms such as $E_\ell^{01} = E_\ell^{10}$
are suppressed.

\begin{figure}[t]
\centerline{\psfig{file=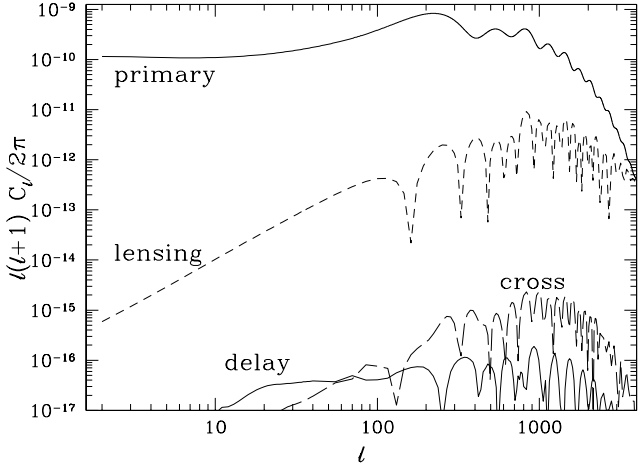,width=3.5in}}
\caption{Delay, lensing and delay-lensing (cross) perturbations to the CMB temperature
power spectrum for the fiducial $\Lambda$CDM model.  The cross spectrum dominates
the delay spectrum but both produce negligible changes to the primary anisotropies
unlike lensing.}
\label{fig:temppower}
\end{figure}

\section{Temperature Power Spectrum}
\label{sec:power}

The perturbations to the power spectrum due to the lensing and time-delay effects
follow by considering the second order
terms in the two-point correlation of the temperature field in equation~(\ref{eqn:expand}).
We can express the contributions schematically as
\begin{equation}
C_\ell^{\Theta\Theta} =
T_\ell^{00} +
(\dC_\ell^{\len\len} + \dC_\ell^{\len^2}) +
(\dC_\ell^{\del\del} + \dC_\ell^{\del^2}) +
\dC_\ell^{\len\del} \,.
\end{equation}
We will now define and consider each term in turn.

\subsection{Lensing Spectra}
\label{sec:geopower}

The pure lensing contributions can be defined as
\begin{eqnarray}
\dC_\ell^{\len \len} & \equiv &
\int {d k \over k} {k^3 \over 2\pi^2} {1 \over 2\ell+1}
\sum_m\left< \Theta^{\len_1 *}_{\ell  m}  \Theta^{\len_1}_{\ell m} \right> \,,
\nonumber\\
\dC_\ell^{\len^2}  & \equiv &
\int {d k \over k} {k^3 \over 2\pi^2} {1 \over 2\ell+1}
\sum_m {1 \over 2}\left(
\left< \Theta^{\prim *}_{\ell  m}  \Theta^{\len^2}_{\ell m} \right> + {\rm cc.}\right)\,.
\end{eqnarray}

Following Ref. \cite{Hu00}, we
expand the
perturbations to the temperature field from equation~(\ref{eqn:pert})
and the projected potential $\phi$ in spherical harmonics  to obtain
\begin{eqnarray}
\dC_\ell^{\len \len}  &=&  \sum_{\ell_1 \ell_2}  C_{\ell_1}^{\len\len} T_{\ell_2}^{00}
				W_{\ell\ell_1\ell_2}^2 L_{\ell\ell_1\ell_2}^2
\,,\nonumber\\
\dC_\ell^{\len^2}  &=&  - {1 \over 2}\theta_\rms^2 \ell(\ell+1) T_\ell^{00}\,,
\label{eqn:lensingpower}
\end{eqnarray}
where
\begin{eqnarray}
W_{\ell\ell_1\ell_2} &=& \sqrt{(2\ell_1+1)(2\ell_2+1) \over 4\pi} \wj \,,
\nonumber\\
L_{\ell\ell_1\ell_2} &=&{1 \over 2} [\ell_1(\ell_1+1)+\ell_2(\ell_2+1)-\ell(\ell+1)] \, .
\end{eqnarray}
The $W_{\ell\ell_1\ell_2}$ term comes from the integral over the product of
three spherical harmonics; the
$L_{\ell\ell_1\ell_2}$ term comes from the conversion of angular
gradients into angular Laplacians through integration by parts.
On scales where there is power in the primary power spectrum,
the two terms in equation~(\ref{eqn:lensingpower}) nearly cancel.  The reason is
that the large-angle modulation produced by lensing simply
smoothes the features in the power spectrum across
$\Delta\ell \sim \ell_{1/2}\approx 30$ \cite{Sel96}.
The result of combining the two terms in the fiducial $\Lambda$CDM model is displayed
in Fig.~\ref{fig:temppower}.

\subsection{Delay Spectra}

The time-delay modifications to the power spectra can be defined as
\begin{eqnarray}
\dC_\ell^{\del \del} & \equiv &
\int {d k \over k} {k^3 \over 2\pi^2} {1 \over 2\ell+1}
 \sum_m \left< \Theta^{\del_1 *}_{\ell  m}  \Theta^{\del_1}_{\ell  m} \right> \,,\nonumber\\
\dC_\ell^{\del^2}    & \equiv &
\int {d k \over k} {k^3 \over 2\pi^2} {1 \over 2\ell+1}
 \sum_m {1 \over 2} \left(
\left< \Theta^{\prim *}_{\ell  m}  \Theta^{\del^2}_{\ell m} \right> + {\rm cc}\right) \,,
\end{eqnarray}
and their evaluation follows closely that of lensing
except for the simplification that no derivatives of spherical harmonics are involved.
The result is
\begin{eqnarray}
\dC_\ell^{\del \del} & = & \sum_{\ell_1 \ell_2}  C_{\ell_1}^{\del\del} T_{\ell_2}^{11}
		W_{\ell\ell_1\ell_2}^2 \,,\nonumber\\
\dC_\ell^{\del^2}  &=&  d_{\rms}^2 T_\ell^{02} \,,
\label{eqn:delaypower}
\end{eqnarray}
where
we have used the identity
\begin{equation}
\sum_{m_1}
Y_{\ell_1}^{m_1*} Y_{\ell_1}^{m_1}
={2 \ell_1 +1 \over 4\pi}\,,
\end{equation}
to evaluate
\begin{equation}
\sum_{m_1} \int d\bn Y_\ell^{m *} Y_{\ell_1}^{m_1*} Y_{\ell_1}^{m_1} Y_{\ell_2}^{m_2}
= {2 \ell_1 +1 \over 4\pi} \delta_{\ell,\ell_2} \delta_{m, m_2}\,.
\end{equation}
The main difference between the lensing and time delay contributions is that
the power spectra of the angular gradients of the projected potential $\ell (\ell+1)
C_\ell^{\len\len}$ and primary temperature anisotropies  $\ell (\ell+1) \dC_\ell^{00}$
are replaced by the power spectra of the delay $C_\ell^{\del\del}$
and that of the {\it radial} derivatives of the temperature field $T_\ell^{11}\approx -T_\ell^{02}$.

As in the case of lensing, the time-delay effect represents a smoothing of
the gradient power spectrum $T_\ell^{11}$ across a width of $\Delta\ell \sim
\ell_{1/2} \approx 2$.  The effect is strongly suppressed by the small width
of the smoothing and the nearly featureless  underlying spectrum $T_\ell^{11}$
(see Fig.~\ref{fig:gradpower}).
Contrast this with lensing which has a smoothing width of $\Delta \ell \sim 30$
and an angular gradient power spectrum that shows peaks from the acoustic
oscillations.   Acoustic features in the power spectrum
arise from features in the source power spectrum
when the source wavevector is oriented perpendicular
to the line of sight.  The source then projects with a one-to-one correspondence
between angular and physical scale ($\ell = k r$ see Fig.~\ref{fig:delay}).
Other alignments contribute to a broad tail to lower multipole moments (see Fig.~\ref{fig:jl}).
In the perpendicular orientation, however,
a perturbation to the radial distance simply
moves the last scattering surface {\it along} the crests and troughs of the source
leaving no net effect.
Mathematically, this effect can be seen in the fact that $j_\ell(kr)$ as a function of
$\ell$ possesses a strong peak at $\ell = kr$, whereas $j_\ell'(kr)$ does not
(see Fig.~\ref{fig:jl}).

The net effect is therefore much smaller than the naive scaling of the lensing
rms~(\ref{eqn:rmslensing}) and delay rms~(\ref{eqn:rmsdelay}) would imply.
It is shown in Fig.~\ref{fig:temppower} for the fiducial model.

\subsection{Cross Spectra}

The cross correlation between the lensing and delay cause modifications defined by
\begin{eqnarray}
\dC_\ell^{\len \del} & \equiv &
\int {d k \over k} {k^3 \over 2\pi^2} {1 \over 2\ell+1} \sum_m \Big(
	 \left< \Theta^{\del_1 *}_{\ell m}  \Theta^{\len_1}_{\ell m} \right>
\nonumber\\
&& + {1 \over 2}\left< \Theta^{\prim *}_{\ell m}  \Theta^{\len \del}_{\ell m}
	\right>  + {\rm cc.} \Big)\,,
\end{eqnarray}
The $\Theta^{\len\del}$ terms are identically zero  since
\begin{eqnarray}
\sum_m [ (\nabla_i Y_{\ell}^{m *})
	 Y_{\ell}^{m}
+
         Y_{\ell}^{m *} (\nabla_i Y_{\ell}^{m})]
&=&
\nabla_i \sum_m (Y_{\ell}^{m *} Y_\ell^{m}) \nonumber\\
&=& 0\,,
\end{eqnarray}
by virtue of the addition theorem of spherical harmonics.
The first term reduces to
\begin{eqnarray}
\dC_\ell^{\len\del} & = &
	 	2	\sum_{\ell_1 \ell_2}  C_{\ell_1}^{\len\del}
	T_{\ell_2}^{01}
				W_{\ell\ell_1\ell_2}^2 L_{\ell\ell_1\ell_2}\,.
\end{eqnarray}
Unlike the pure lensing and time delay contributions, there is no
second canceling term.
However as discussed in \S \ref{sec:temp}, $T_\ell^{01}$ is intrinsically
small reflecting the lack of correlation between $j_\ell$ and $j_\ell'$ (see
Fig.~\ref{fig:jl}).
Nonetheless, its net contribution to temperature anisotropy power spectrum
is still larger than the pure delay contribution, as a result of
the larger amplitude and smaller
coherence of the lensing-delay power spectrum $C_\ell^{\len\del}$
(see Fig.~\ref{fig:temppower}).

\begin{figure}[t]
\centerline{\psfig{file=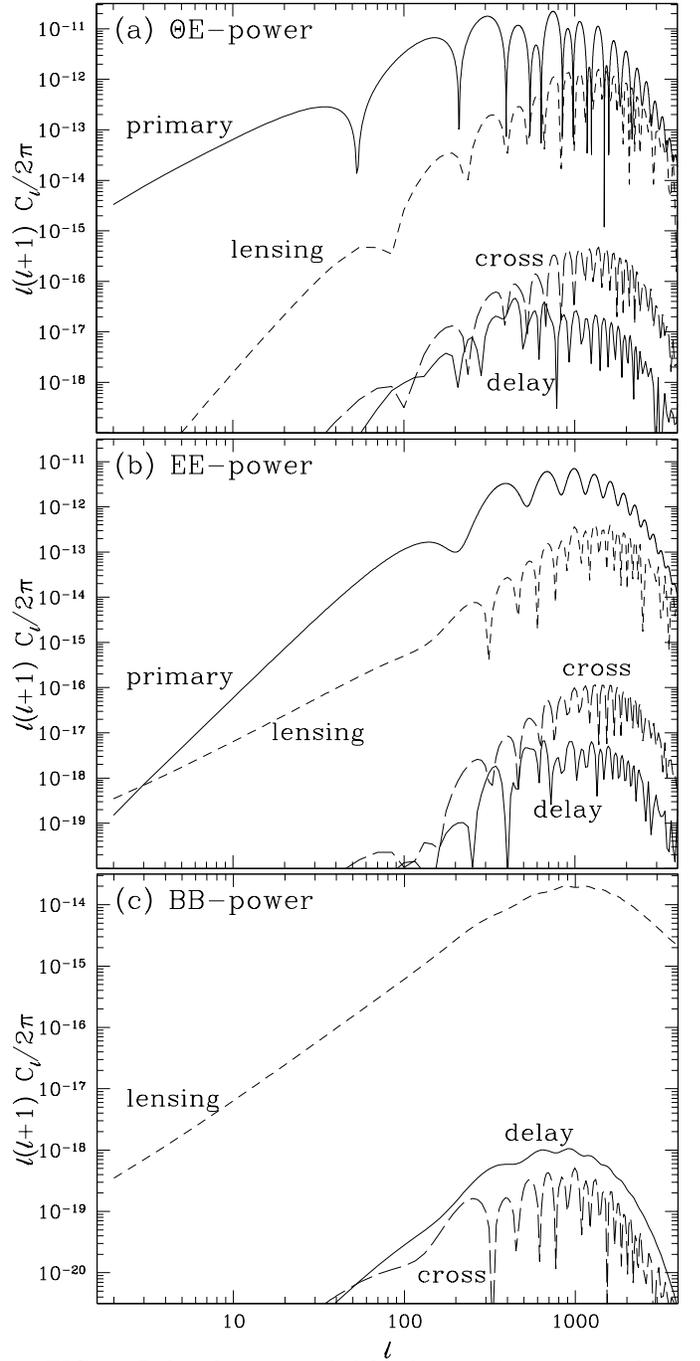,width=3.5in}}
\caption{Delay, lensing and delay-lensing (cross) contributions to
the CMB polarization power spectra for the fiducial $\Lambda$CDM model.
The delay-lensing cross contributions to the temperature-polarization
power spectrum reaches $10^{-3}$ of the primary and/or lensing power.
Delay effects only weakly generate power in the $BB$ spectrum.}
\label{fig:polpower}
\end{figure}

\section{Polarization Power Spectra}
\label{sec:polarization}

Evaluation of the lensing and delay effects on
the polarization and temperature-polarization cross
power spectra follows precisely the same steps as that of the temperature power spectrum
considered above.
We can express the contributions schematically as
\begin{eqnarray}
C_\ell^{EE} &= & \dE_\ell^{00} +
(\dE_\ell^{\len\len} + \dE_\ell^{\len^2}) +
(\dE_\ell^{\del\del} + \dE_\ell^{\del^2}) +
\dE_\ell^{\len\del} \,, \nonumber\\
C_\ell^{BB} &= & \dB_\ell^{00} +
(\dB_\ell^{\len\len} + \dB_\ell^{\len^2}) +
(\dB_\ell^{\del\del} + \dB_\ell^{\del^2}) +
\dB_\ell^{\len\del} \,, \\
C_\ell^{\Theta E} &= & \dX_\ell^{00} +
(\dX_\ell^{\len\len} + \dX_\ell^{\len^2}) +
(\dX_\ell^{\del\del} + \dX_\ell^{\del^2}) +
\dX_\ell^{\len\del} \,. \nonumber
\end{eqnarray}
We will now consider each term in turn.

\subsection{Lensing Spectra}

Following \cite{Hu00}, the lensing contributions to the polarization
power spectra can be evaluated as
\begin{eqnarray}
\dE_\ell^{\len\len} &=&
			\sum_{\ell_1\ell_2} C_{\ell_1}^{\len\len}
			{}_+ \Sigma_{\ell_2}^{00}(\ell+\ell_1+\ell_2)
			V_{\ell \ell_1 \ell_2}^2 L_{\ell \ell_1 \ell_2}^2 \,,\nonumber\\
\dE_\ell^{\len^2} &=& -{1 \over 2} \theta_\rms^2 ( \ell^2 + \ell -4) \dE_\ell^{00} \,,\\
&&\nonumber\\
\dB_\ell^{\len\len} &=&
			\sum_{\ell_1\ell_2} C_{\ell_1}^{\len\len}
			{}_- \Sigma_{\ell_2}^{00} (\ell+\ell_1+\ell_2)
			V_{\ell \ell_1 \ell_2}^2 L_{\ell \ell_1 \ell_2} \,,\nonumber\\
\dB_\ell^{\len^2} &=& -{1 \over 2} \theta_\rms^2 (\ell^2 + \ell -4) \dB_\ell \,,\\
&&\nonumber\\
\dX_\ell^{\len\len} &=&
			\sum_{\ell_1\ell_2} C_{\ell_1}^{\len\len}
			\dX_{\ell_2}^{00}
			W_{\ell \ell_1 \ell_2} V_{\ell \ell_1 \ell_2} L_{\ell \ell_1 \ell_2}^2
			\nonumber\,,\\
\dX_\ell^{\len^2} &=& -{1 \over 2} \theta_\rms^2 ( \ell^2 + \ell -2) \dX_\ell \,,
\end{eqnarray}
where
\begin{eqnarray}
{}_\pm \Sigma_\ell^{ab} (L) =
				{[1+(-1)^L] \over 2}
				{\dE}_\ell^{ab} \pm
				{[1-(-1)^L] \over 2}
				{\dB}_\ell^{ab}\,.
\end{eqnarray}
Here
\begin{equation}
V_{\ell \ell_1 \ell_2} = \sqrt{(2 \ell_1+1) (2 \ell_2 +1) \over 4\pi}
\wjma{\ell}{\ell_1}{\ell_2}{2}{0}{-2}  \,,
\end{equation}
comes from the integral over the product of a spherical harmonic with
2 spin-$2$ spherical harmonics.
As in the case of the temperature power spectra, the main effect on
the $\Theta E$ and $EE$ power spectra is a smoothing by $\Delta\ell \sim
\ell_{1/2}$.  If there were an intrinsic $BB$ power spectrum, the
smoothing would also apply.   However,
since scalar perturbations do not generate $BB$ modes
in the primary polarization ($\dB_\ell^{00}=0$),  the
generation of $B$-polarization from
the lensing modulation of the primary $E$-polarization dominates
\cite{ZalSel98}.
The amount of
generation is again related to the coherence scale $\ell_{1/2} \approx \ell_1$
of the effect as reflected
in the difference between even and odd $L$ terms in $V_{\ell \ell_1 \ell_2}$.
Considering the triplet as forming a triangle, even terms are associated with
the cosine of twice the opening angle
between $\ell$ and $\ell_2$; odd terms are
associated with the sine of that angle (\cite{Hu00}, Eqns. (B8) and (B10)).
These polarization contributions are
shown in Fig.~\ref{fig:polpower} for the fiducial model.

\subsection{Delay Spectra}
\label{sec:tdpower}

The derivation of the delay power spectra follows the same
steps as those involved in the lensing derivation yielding
\begin{eqnarray}
\dE_\ell^{\del\del} &=&
			\sum_{\ell_1\ell_2} C_{\ell_1}^{\del\del}
	{}_+ \Sigma_{\ell_2}^{11}(\ell+\ell_1+\ell_2)
			V_{\ell \ell_1 \ell_2}^2
			\,,\nonumber\\
\dE_\ell^{\del^2} &=& d_\rms^2 E_{\ell}^{02} \,,\\
&&\nonumber\\
\dB_\ell^{\del\del} &=&
			\sum_{\ell_1\ell_2} C_{\ell_1}^{\del\del}
	{}_- \Sigma_{\ell_2}^{11}(\ell+\ell_1+\ell_2)
			V_{\ell \ell_1 \ell_2}^2 \,, \nonumber\\
\dB_\ell^{\del^2} &=& d_\rms^2 B_{\ell}^{02} \,, \\
&&\nonumber\\
\dX_\ell^{\del\del} &=&
			\sum_{\ell_1\ell_2} C_{\ell_1}^{\del\del}
			X_{\ell_2}^{11}
			W_{\ell \ell_1 \ell_2} V_{\ell \ell_1 \ell_2}  \,,\nonumber\\
\dX_\ell^{\del^2} &=& d_\rms^2 X_{\ell}^{02} \,.
\end{eqnarray}
The results of summing these nearly canceling pairs for the fiducial model
are shown in Fig.~\ref{fig:polpower}.  A $BB$-spectrum is generated out of
the primary $EE$-spectrum but at an efficiency that is substantially below
that of gravitational lensing.  The underlying reason again is that the coherence
of the effect $\ell_{1/2} \sim 2$.

\subsection{Cross Spectra}
\label{sec:crosspower}

The cross spectra contributions are defined as
\begin{eqnarray}
\dE_\ell^{\phi\del} &=& 2\sum_{\ell_1 \ell_2}  C_{\ell_1}^{\phi d}
	{}_+ \Sigma_{\ell_2}^{01} (\ell+\ell_1+\ell_2)
		V_{\ell \ell_1 \ell_2}^2
		L_{\ell \ell_1 \ell_2} \,,
\nonumber\\
\dB_\ell^{\phi\del} &=& 2\sum_{\ell_1 \ell_2}  C_{\ell_1}^{\phi d}
	{}_- \Sigma_{\ell_2}^{01} (\ell+\ell_1+\ell_2)
		V_{\ell \ell_1 \ell_2}^2
		L_{\ell \ell_1 \ell_2} \,,\nonumber\\
\dX_\ell^{\phi\del} &=& \sum_{\ell_1 \ell_2}  C_{\ell_1}^{\phi d}
		        (X_{\ell_2}^{01}+X_{\ell_2}^{10})
  	        W_{\ell \ell_1 \ell_2}
		V_{\ell \ell_1 \ell_2}
		L_{\ell \ell_1 \ell_2} \,.
\end{eqnarray}
Figure \ref{fig:polpower} shows that the cross spectra dominate over the pure delay
spectra for the $\Theta E$ and $EE$ power spectra.  In particular, since the
lensing effects themselves approach order unity at $\ell \sim 1000$, the lensing-delay
cross effects reach $\sim 10^{-3}$ of the primary $C_\ell^{\Theta E}$ power spectrum.
While still small, the contribution is of order the cosmic variance out to comparable
multipoles.

\begin{figure}[t]
\centerline{\psfig{file=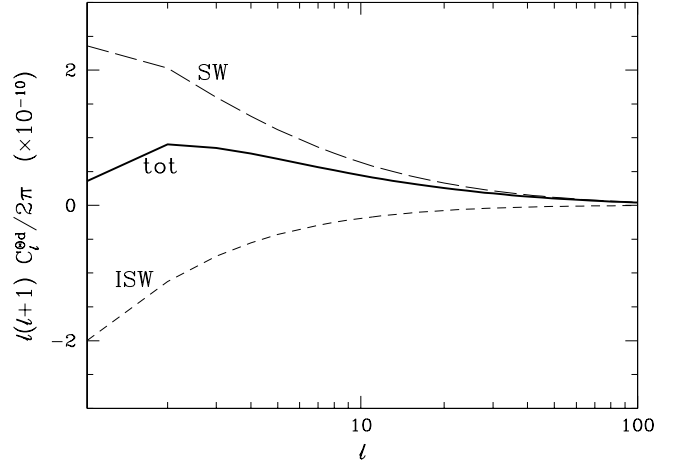,width=3.5in}}
\caption{Delay-temperature cross correlation.  The cross-correlation between the
temperature and delay fields, as relevant for the bispectra, are dominated by contributions
from the Sachs-Wolfe (SW) effect with partially canceling contributions from the
integrated Sachs-Wolfe effect (ISW).}
\label{fig:bicross}
\end{figure}

\section{Delay Bispectrum}
\label{sec:bispectrum}

Second order effects generally produce non-Gaus\-sianity in the
CMB temperature and polarization fields.   Effects that provide
a negligible change to the power spectrum can in principle produce
observable effects due to the expected Gaussianity of the primary
anisotropies.  Here we consider the three-point correlations induced
by time-delay in angular harmonic space, i.e. the bispectrum.

The angle averaged bispectrum is defined as
\begin{equation}
B_{\ell \ell' \ell''}^{F F' F''} = \sum_{m m' m''}
\wjma{\ell}{\ell'}{\ell''}{m}{m'}{m''}
\left< F_{\ell m} F_{\ell' m'} F_{\ell'' m''} \right>\,,
\end{equation}
where the $F$'s can take on the values $\Theta$, $E$, $B$ for the
temperature and polarization components respectively.
Since the derivation follows closely that of the lensing
bispectra terms, we refer the reader to \cite{Hu00} for detailed
derivations.

\begin{figure}[t]
\centerline{\psfig{file=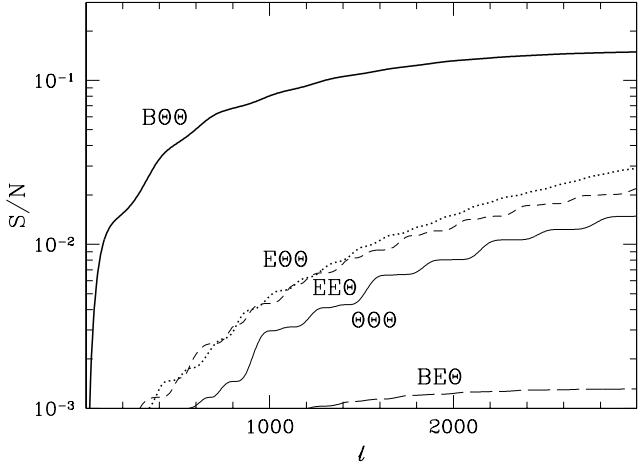,width=3.5in}}
\caption{Signal-to-noise for detection of bispectra involving gravitational
time-delay in an ideal cosmic-variance limited experiment.}
\label{fig:sn}
\end{figure}

\subsection{Temperature}

The temperature bispectrum
produced by time delays follows immediately from
the decomposition in equation~(\ref{eqn:pert})
\begin{equation}
B_{\ell_1 \ell_2 \ell_3}^{\Theta\Theta\Theta}
= \sqrt{2\ell_1+1} W_{\ell_1 \ell_2 \ell_3} C_{\ell_2}^{\Theta d}
	T_{\ell_3}^{0 1} + {\rm 5 perm.}\,,
\end{equation}
where the permutations are with respect to the $\ell$-indices
and
\begin{equation}
\left< \Theta_{\ell m}^* d_{\ell' m'} \right> =
\delta_{\ell \ell'} \delta_{m m'} C_\ell^{\Theta d} \,.
\end{equation}
Unlike lensing however, $C_\ell^{\Theta d}$ is
not solely the result of correlations with secondary anisotropies
but includes relatively large contributions from the
primary anisotropies themselves.  The delay field
arises from potential fluctuations of sufficiently large scale that
it is correlated with the Sachs-Wolfe effect \cite{SacWol67} in the primary
anisotropies.  To see this, let us approximate the primary anisotropies
at large angles as the sum of Sachs-Wolfe (SW) and integrated Sachs-Wolfe
(ISW) contributions
\begin{equation}
\Theta(\bn) \approx -{1 \over 3} \Phi(r_*\bn,\eta_*)
	- \int d \eta'\, 2\dot\Phi(r'\bn,\eta')   \,,
\end{equation}
where the asterisk denotes evaluation at recombination.
This temperature field may be directly correlated with the delay field in
equation~(\ref{eqn:projectedtd}).  The resulting power spectrum
is shown in Fig.~\ref{fig:bicross}.  Note that the SW and ISW
contributions cancel at the lowest multipoles.  Unlike the
lensing-ISW correlation \cite{sec}, the delay effects on the bispectrum
do not vanish and indeed increase as $\Omega_\Lambda \rightarrow 0$.

To determine whether the contributions are detectable, we evaluate
the signal-to-noise ratio for an ideal cosmic variance limited
experiment
\begin{equation}
\left( { S \over N} \right)_{\Theta\Theta\Theta}^2 =
	\sum_{\ell_1 \ell_2 \ell_3} {
		(B_{\ell_1 \ell_2 \ell_3}^{\Theta\Theta\Theta})^2
	\over 6 C_{\ell_1}^{\Theta\Theta} C_{\ell_2}^{\Theta\Theta}
		C_{\ell_3}^{\Theta\Theta} } \,.
\end{equation}
We show the cumulative signal-to-noise out to a given maximum
$\ell$ in Fig.~\ref{fig:sn}.  The time-delay temperature
bispectrum is not detectable even for an ideal experiment.
This should be compared with the lensing-temperature bispectrum, where
the cross-correlation between lensing and effects such
as SZ can be detected with upcoming experiments, such as Planck.

\subsection{Polarization}

Bispectrum terms involving the polarization follow similarly.  The
non-vanishing contributions are
\begin{eqnarray}
B^{E\Theta\Theta}_{\ell \ell_1 \ell_2} &=&
	\sqrt{2\ell_1+1} V_{\ell_1 \ell_2 \ell_3}
		C_{\ell_2}^{\Theta d} X_{\ell_3}^{01} \\
 	&& \quad +
	\sqrt{2\ell_2+1} W_{\ell_2 \ell_1 \ell_3}
		X_{\ell_1}^{10} C_{\ell_3}^{\Theta d}
		+ (\ell_2 \leftrightarrow \ell_3)\,,\nonumber\\
B^{E E \Theta}_{\ell \ell_1 \ell_2}   & = &
	\sqrt{2\ell_1+1} V_{\ell_1 \ell_3 \ell_2}
		E_{\ell_2}^{1 0} C_{\ell_3}^{\Theta d}
		+ (\ell_1 \leftrightarrow \ell_2)\,,
\end{eqnarray}
for
$\ell_1 + \ell_2 + \ell_3 =$ even
and
\begin{eqnarray}
B^{B \Theta\Theta}_{\ell \ell_1 \ell_2} &=& i
	\sqrt{2\ell_1+1} V_{\ell_1 \ell_2 \ell_3}
		C_{\ell_2}^{\Theta d} X_{\ell_3}^{01}
		+ (\ell_2 \leftrightarrow \ell_3)\,,\\
B^{B E \Theta}_{\ell \ell_1 \ell_2} & = & i
	\sqrt{2\ell_1+1} V_{\ell_1 \ell_2 \ell_3}
		E_{\ell_2}^{10} C_{\ell_3}^{\Theta d}\,,
\end{eqnarray}
for
$\ell_1 + \ell_2 + \ell_3 =$ odd.  The signal-to-noise ratio in
an ideal, cosmic variance limited experiment is
\begin{eqnarray}
\left( { S \over N} \right)_{E\Theta\Theta}^2 & \approx &
	\sum_{\ell_1 \ell_2 \ell_3} {
		(B_{\ell_1 \ell_2 \ell_3}^{E\Theta\Theta})^2
	\over 6 C_{\ell_1}^{E E} C_{\ell_2}^{\Theta\Theta}
		C_{\ell_3}^{\Theta\Theta} } \,, \\
\left( { S \over N} \right)_{E E\Theta}^2 & \approx &
	\sum_{\ell_1 \ell_2 \ell_3} {
		(B_{\ell_1 \ell_2 \ell_3}^{E E \Theta})^2
	\over 6 C_{\ell_1}^{E E} C_{\ell_2}^{EE}
		C_{\ell_3}^{\Theta\Theta} } \,, \\
\left( { S \over N} \right)_{B \Theta\Theta}^2 & \approx &
	\sum_{\ell_1 \ell_2 \ell_3} {
		(B_{\ell_1 \ell_2 \ell_3}^{B\Theta\Theta})^2
	\over 2 C_{\ell_1}^{B B} C_{\ell_2}^{\Theta \Theta}
		C_{\ell_3}^{\Theta\Theta} } \,, \\
\left( { S \over N} \right)_{B E\Theta}^2 & \approx &
	\sum_{\ell_1 \ell_2 \ell_3} {
		(B_{\ell_1 \ell_2 \ell_3}^{B E\Theta})^2
	\over 2 C_{\ell_1}^{B B} C_{\ell_2}^{EE}
		C_{\ell_3}^{\Theta\Theta} } \,.
\end{eqnarray}
In Fig.~\ref{fig:sn}, we show the cumulative signal-to-noise
ratio
as a function of the maximum $\ell$.  The signal-to-noise ratio for
the $B\Theta\Theta$ term is larger than the others since
we have assumed that $C_\ell^{BB}$ vanishes for the primary
anisotropies and is only generated by the lensing and delay
effects.  Nonetheless, the signal-to-noise is substantially
less than unity and so the time-delay effects are unlikely
to be detectable in the bispectrum or affect the extraction of
other effects from the bispectrum.

\section{Discussion}
\label{sec:discussion}

Gravitational time delays introduce a radial perturbation in the
mapping of the CMB temperature field at recombination
onto temperature and polarization
anisotropies today.  The effect is closely related to gravitational
lensing which introduces angular perturbations in the same mapping.
Despite the fact that radial perturbations are only one order
of magnitude smaller than angular perturbations, and moreover highly correlated with
the angular perturbations, their effect on the power spectra and bispectra
of the temperature and polarization fields is approximately 3 orders of magnitudes smaller.
The underlying reason is that on the angular scales of the acoustic
peaks neither effect actually generates new anisotropies; both
induce a large scale modulation of the primary field.  Angular modulations
produce a substantial effect due to angular structure in the
acoustic peaks by smoothing the power spectrum on the coherence
scale $\Delta \ell \sim 30$. Radial modulations produce
much smaller effects due to the lack of radial structure
in the perturbations that form the acoustic peaks.  
They also suffer from the fact that the angular coherence 
or smoothing scale of the radial modulation is typically one 
order of magnitude larger $\Delta \ell \sim 2$.

As a result, at $\ell \sim 1000$, the delay and lensing-delay
correlation effects are $\simlt 10^{-4}$ of the temperature,
$E$-polarization and lensing-induced $B$-polarization power spectrum
generated by lensing.  For the temperature-polarization cross
correlation, the lensing-delay correlation effect approaches
$10^{-3}$ and so is comparable to the cosmic variance on these scales.
This enhancement reflects the same efficiency with which lensing
modulations affect the temperature-polarization correlation.
It is therefore relevant in principle for the Planck satellite or any
future experiment that expects to be
cosmic variance limited at $\ell \simgt 1000$.
In practice, there may well be other more limiting sources of systematic
errors such as galactic and extragalactic foregrounds.

For the bispectra, the time delay couples
mainly with the Sachs-Wolfe effect in the primary anisotropies.
For a cosmic variance limited experiment, the signal-to-noise is
highest in the $B$-temperature-temperature bispectrum
since the $B$-polarization
vanishes for primary anisotropies from scalar perturbations.
In a realistic experiment, the low level of the signal will make
the cosmic variance limit difficult to achieve.  In any case,
the signal-to-noise ratios in the bispectra never exceed the $10^{-1}$
level for $\ell \sim 1000$ and hence are unlikely to interfere with the extraction
of signals in the bispectrum from secondary anisotropies by
the next generation of satellites \cite{sec}.

The potential of cosmic microwave background anisotropies for
studying cosmology is considered vast
primarily because of the physical processes underlying their
formation are thought to be understood to extraordinary precision
relative to other astrophysical systems.  Though unlikely to
affect the next generation of experiments, small effects such
as the gravitational time delay considered here
must be calculated and included to ensure that this potential is realized.

\acknowledgments
We thank Sean Carroll and Matias Zaldarriaga for useful discussions.
WH is supported by the Keck Foundation, a Sloan Fellowship,
and NSF-9513835.
ARC acknowledges financial support from Don York and computational
support from John Carlstrom.
We acknowledge the use of CMBFAST \cite{SelZal96}
and the routine to generate spherical Bessel functions
from Arthur Kosowsky.

\end{document}